\documentclass[prb,aps,twocolumn, longbibliography,floatfix]{revtex4-2}
\usepackage[
  colorlinks=true,
  urlcolor=blue,
  linkcolor=blue,
  citecolor=blue,
  bookmarks=false,
  pdftitle={},
  pdfauthor={}
]{hyperref}
\usepackage{import}
\usepackage{physics}
\usepackage{subfiles}
\usepackage{graphicx}
\usepackage{bm}
\usepackage{amsmath}
\usepackage{amssymb}
\usepackage{mathtools}
\usepackage{amsfonts}
\usepackage[all]{xy}
\usepackage{xspace}
\usepackage{accents}
\usepackage{stmaryrd}
\usepackage{tabularx}
\usepackage{extarrows}
\usepackage{float}
\usepackage{multirow}
\usepackage{makecell}
\usepackage[capitalise]{cleveref}
\usepackage[dvipsnames]{xcolor}
\usepackage[normalem]{ulem}
\usepackage{soul}
\usepackage{pdfpages}
\makeatletter
\AtBeginDocument{\let\LS@rot\@undefined}
\makeatother

\begin{document}

\title{Breakdown of chiral anomaly and emergent phases in Weyl semimetals
under orbital magnetic fields}
\author{Faruk Abdulla}
\author{Anna Keselman}
\author{Daniel Podolsky}
\affiliation{Physics Department, Technion - Israel Institute of 
Technology, Haifa 32000, Israel}
\affiliation{The Helen Diller Quantum Center, Technion, Haifa 32000, Israel}

\begin{abstract}

An external orbital magnetic field applied perpendicular to the separation 
vector of a pair of Weyl points can couple them and induce a gap in the 
electronic spectrum. In this work, we investigate the gap-opening behavior in 
the presence of a lattice, revealing rich phenomenology absent in the continuum 
picture. Specifically, we address the emergence of layered Chern insulating 
states, examining how the anisotropy of the Weyl cone dispersion influences 
the sequence of phase transitions, and establishing connections to the continuum 
limit. We analyze the evolution of surface Fermi-arc states across these 
regimes, highlighting their distinct behaviors during the gap-opening transitions.

\end{abstract}

\maketitle

\section{Introduction}

Weyl semimetals (WSMs) \cite{Murakami_2007, Wan_Savrasov_2011,Lv_Ding_2015a, 
Lv_Ding_2015b, Xu_Hasan_2015a, Xu_Hasan_2015b, Lu_Soljacic_2015} 
represent a robust gapless topological phase of 
quantum matter in three dimensions. As long as spatial translation 
symmetry is maintained, the only way to eliminate this phase is 
by bringing pairs of Weyl nodes (WNs) with opposite chiralities 
together, allowing them to annihilate each other. WSMs exhibit a 
variety of exotic phenomena -- the chiral anomaly 
\cite{Nielsen_Ninomiya_1983, Bevan1997, Aji_2012, Zyuzin_Burkov_2012}, 
chiral anomaly-induced negative magnetoresistance \cite{Son_Spivak_2013, 
Gorbar_Miransky_2014, Burkov_2015, Das_Axialanomaly_2015, Huang_2015, 
Zhang_Hasan_2016,Li_Das_2016, Lu_Shun_2017}, planar Hall 
effect \cite{Nandy_Tewari_2017, Li_Shen_2018, 
Shama_Singh_2020, Li_Yao_2023, Wei_Weng_2023}, Fermi-arc-mediated quantum 
oscillations, and three-dimensional quantum Hall effect \cite{Potter_Vishwanath_2014, Zhang_Vishwanath_2016,
Moll_Analytis_2016, Wang_Xie_2017, Zhang_Xiu_2019, Li_Xie_2020, 
Chang_Xing_2021,  Ma_Sheng_2021, Chang_Sheng_2022, Zhang_Naoto_2022} -- 
all of which typically rely on the presence of an external orbital magnetic field. 
However, when the magnetic field strength is sufficiently high, it can couple
WNs of opposite chiralities and open a finite gap in the spectrum, thereby
destabilizing the WSM state \cite{Goerbig_2009, Zhang_Jia_2017, Ramshaw_McDonald_2018, 
Chan_Lee_2017, Kim_Park_2017, Saykin_Rodinov_2018, Bednik_Syzranov_2020, Abdulla_Murthy_2022, Abdulla_2024}.

An external magnetic field that is not aligned with the separation direction
of WNs can couple nodes of opposite chirality, resulting in the opening of a 
gap in the bulk electronic spectrum. This phenomenon has been the subject of 
several theoretical investigations based on low-energy continuum models. 
Notably, Refs.~\onlinecite{Saykin_Rodinov_2018, Bednik_Syzranov_2020} demonstrated 
that the application of an orbital magnetic field immediately induces a gap, 
which is exponentially small, scaling as  $\sim \exp[-(Ql_B)^2]$, where 
$Q$ represents the momentum-space separation between the Weyl nodes and 
$l_B= \sqrt{\hbar/(eB)}$ is the magnetic length associated with the applied 
field. In these models, the gap increases monotonically as the magnetic field 
strength is raised. Furthermore, Ref.~\onlinecite{Devakul_Parameswaran_2021} 
revealed that in systems where the Weyl fermion velocities exhibit anisotropy, 
the induced gap can display oscillatory behavior as a function of the magnetic 
field strength, adding further complexity to the field-induced modification 
of the electronic structure.

It is important to note that these results, derived from continuum approximations, 
are strictly valid within a specific regime of parameters: when the separation
between the Weyl nodes ($Q$) is much smaller than the linear dimensions of the
Brillouin zone (BZ), and when the magnetic length $l_B$ is considerably larger
than the lattice constant $a$, so that lattice effects can be safely ignored. 
However, these conditions are not always
satisfied, particularly under strong magnetic fields where $l_B$ approaches
the lattice scale, as in moir\'e systems, or in materials with relatively large 
node separations. Similar effects can also arise in two and three 
dimensional Dirac materials \cite{Wehling2014} with two or more Dirac 
cones in the spectrum, e.g. graphene \cite{Castro_2009, Katsnelson2007}, and
surfaces of three dimensional topological insulators \cite{Hasan2010, Qi2011}.
In such cases, the low-energy continuum description breaks down, failing to
capture crucial lattice-specific phenomena.

Beyond the bulk electronic gap, another significant limitation of the
continuum framework is its inability to address the behavior of surface states, 
specifically the Fermi-arc states, under the influence of magnetic fields. 
The fate of these topologically protected surface states in the presence of
bulk gap openings remains an open and critical question, particularly in
understanding the complete topological character of the field-induced 
gapped phases. Determining whether the gapped state retains a nontrivial
topological classification or transitions into a trivial insulating state
requires a full lattice-based treatment, beyond the scope of continuum
models. Therefore, it is essential to investigate these effects using
lattice models capable of capturing both bulk and surface phenomena, 
as well as the detailed topological structure of the resulting gapped phases.

\begin{figure}
\includegraphics[width=0.7\linewidth]{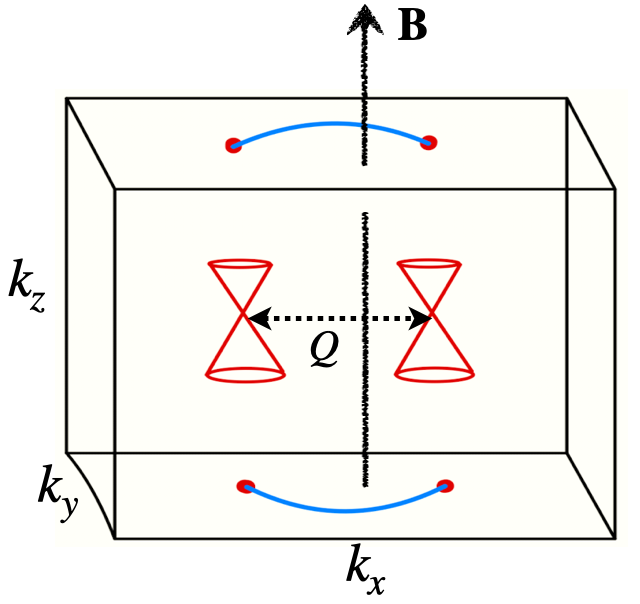}
\caption{Schematic of a Weyl semimetl with two Weyl cones placed in 
a uniform magnetic field ${\bf B} \parallel \hat{z}$. Weyl cones are separated 
along $k_x$ and their momentum space separation is $Q$. Magnetic 
field is aligned perpendicular to the direction of separation between 
Weyl cones of opposite chirality. The orbital field ${\bf B}$ introduces 
a quantum tunneling of particles between the Weyl cones which can induce 
a finite gap in the spectrum. A schematic phase diagram  is depicted 
in Fig. \ref{Fig:Phase_Schematic}. }
\label{Fig:Setup_cartoon}
\end{figure}

In this work, we study magnetic field-induced gaps in WSMs in a lattice.  
We consider the setup depicted in Fig. \ref{Fig:Setup_cartoon}, in which 
two Weyl cones of opposite chirality are separated by a momentum space distance 
$Q$. We assume that both Weyl nodes are at the same energy, which we set to 
be the Fermi energy. For simplicity, we assume zero tilt in the Weyl cones. 
A uniform orbital magnetic field is applied in a direction perpendicular to 
the separation between the Weyl cones.

We find that lattice effects give rise to two important corrections to 
the previous, continuum-limit, results. First, the discrete translational 
symmetry of the lattice implies that $k$-space is periodic i.e. 
${\bf k} \in $ BZ. Therefore, there are always two separations: Intra-BZ 
separation which we have already denoted as $Q$ and inter-BZ separation 
$\vec{Q}'= \vec{G} - \vec{Q}$ ($\vec{G}$ is a reciprocal lattice vector)
between two WNs in a lattice. Consequently, there exist two distinct
ways of gap opening induced by the orbital magnetic fields and hence we expect
two topologically distinct gapped states. Second, the Landau levels (LLs) are 
not perfectly flat. They acquire a finite dispersion which is significant in 
the Hofstadter regime $l_B \sim a$. These dispersive 
LL bands can alter the gap structure in the system. The objective of this 
paper is to provide a comprehensive understanding
of these phenomena. We will initially review the continuum description, 
followed by a detailed investigation of lattice effects.

The paper is organized as follows. In \cref{Sec:Results} we present a summary 
of our main results. In \cref{Sec:Continuum}, we 
revisit the low-energy continuum description of gap opening induced by an 
orbital magnetic field. As we mentioned before, Ref.~\onlinecite{Saykin_Rodinov_2018} 
found the gap to be monotonic with the field, whereas 
Ref.~\onlinecite{Devakul_Parameswaran_2021} found the gap can exhibit periodic oscillations.
In this section, we provide a unified description of the phenomena to set the 
stage for the richer phenomenology in a lattice description. 
The analysis in the presence of a lattice is carried out in \cref{Sec:Lattice}, which 
contains the main results of this work. We  conclude in \cref{Sec:DC}.

\section{Model and Summary of the results}
\label{Sec:Results}

We consider a minimal two-band model of WSM with two WNs which breaks 
time-reversal but preserves inversion.  We assume a cubic lattice model 
with the WNs located on the $k_x$ axis at $k_x = \pm k_0$, with the Hamiltonian 
\begin{equation} \label{Eq:LatWsm}
H({\bf k}) =  2\tilde{v}_x m({\bf k}) \sigma_x +  2v_y\sin{(k_ya)} \sigma_y 
+ 2v_z \sin{(k_za)}\sigma_z, 
\end{equation} 
where $m({\bf k}) = 2t^2 + (\cos{k_0a} - \cos{k_xa}) - t^2(\cos{k_ya} + \cos{k_za})$, $\tilde{v}_x=v_x/\sin{k_0a}$, and $a$ is the lattice constant (we will often work 
in units in which $a=1$).  Here, $\sigma_{x,y,z}$ are Pauli matrices. The parameter 
$k_0$, which sets a separation of $Q=2k_0$ between the Weyl cones, lies in the 
range $0 < k_0 < \pi/a$. In general, the term $m_0({\bf k}) \sigma_0$ is
allowed in the above Hamiltonian (inversion 
symmetry $P=\sigma_x$ constraints $m_0({\bf k})$ to be an even function of momentum). 
It introduces a tilt to the Weyl cone and
also shifts the WNs in energy and can have significant effect on the magnetic 
field induced gaps. In what follows we will ignore this term and focus on the 
relatively simple but still phenomenologically rich situation described by 
the Hamiltonian in Eq.~\eqref{Eq:LatWsm}. In the following, we will assume magnetic 
field is aligned along the $z$ direction and work in the Landau gauge ${\bf A} 
= (-y, 0, 0)B$, unless explicitly stated otherwise.

\begin{figure*}
\includegraphics[width=0.9\linewidth, height=0.45\linewidth]{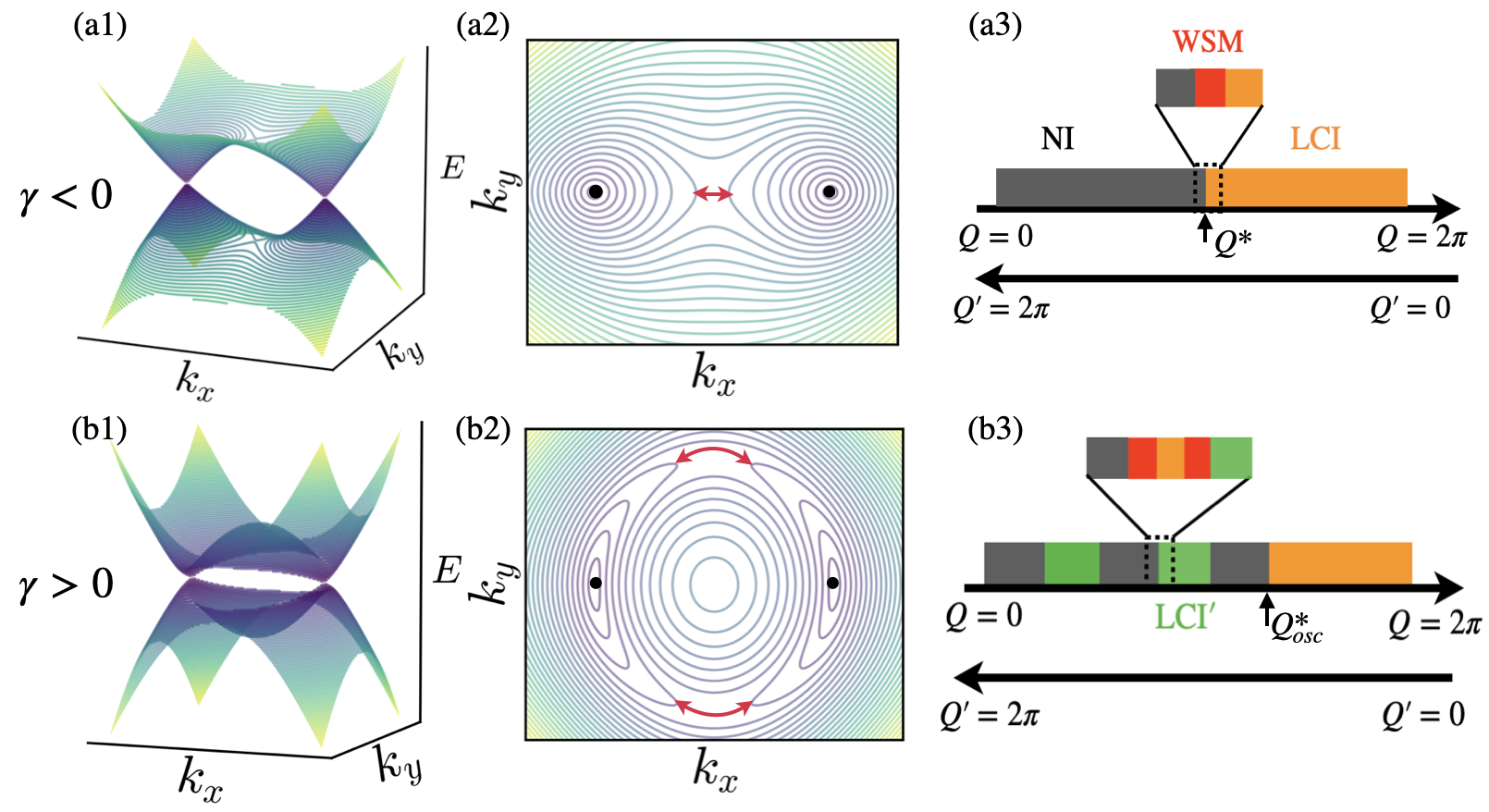}
\caption{Band structures for $\gamma<0$ and $\gamma>0$ are shown in 
(a1) and (b1) respectively. (a2)-(b2) Constant energy contours around 
the Weyl nodes have elliptical and crescent shape for $\gamma<0$ and 
$\gamma>0$ respectively.  Weyl nodes are 
represented by black dots. There is only a single close encounter for 
$\gamma<0$, and two close encounters for $\gamma>0$ between the 
Fermi pockets around the Weyl nodes. 
(a3)-(b3) Schematic phase diagram (for a system on a lattice) in presence of an orbital magnetic  field ${\bf B} \parallel \hat{z}$ which is perpendicular to the direction of separation between two Weyl nodes of opposite chirality. 
Here $Q$ and $Q'$ represent the intra and inter BZ 
separation between the Weyl nodes in momentum space.
(a3) For $\gamma<0$, the system displays three phases as a function of the separation between the Weyl nodes $Q$. The transition from normal insulator
(NI) to layered Chern insulator (LCI) passes through a gapless WSM phase. The width of the WSM region is exponentially small in $\phi_B^{-1}$. (b3) For $\gamma>0$ and a fixed applied flux, there is a series of phase transitions as we tune the separation $Q \in(0, 2\pi)$. Besides NI and LCI, another insulating state which we denote as LCI$'$ appears. The state LCI$'$ can be viewed as two copies of LCI with opposite Chern numbers. All transitions between NI and LCI$'$ phases pass through a sequence of three phases: two WSMs and an LCI. }
\label{Fig:Phase_Schematic}
\end{figure*}

The setup is depicted in 
Figure \ref{Fig:Setup_cartoon}. For a field along the $z$ 
direction, the momentum along $\hat{z}$ remains conserved and we can 
set it to $k_z=0$, since our focus is on low-energy properties. The key parameters 
in the problem are the momentum-space separation between the Weyl cones, $Q$, 
the velocity components $(v_x, v_y)$ of the Weyl fermion perpendicular to the
magnetic field and the parameter $t$.  From these we define a single parameter 
\begin{align}\label{Eq:Anisotropy_pm}
  \gamma = \left(1 - \frac{1}{t^2} \frac{v_y^2}{v_x^2}\right) 
\end{align}
which captures the anisotropy of the Weyl cone dispersion and controls the distinct 
behaviors of the induced gap.  This parameter will play a  
crucial role in the subsequent analysis.  The sign of $\gamma$ determines the shape of the Fermi pockets around the Weyl nodes, which are elliptical for $\gamma<0$ and crescent-shaped for $\gamma>0$, as shown in Fig.~\ref{Fig:Phase_Schematic}.

We find that an orbital magnetic field perpendicular to the direction of separation between 
Weyl cones introduces a tunneling of electrons between them, thus inducing a 
gap in the electronic spectrum. The gap is non perturbative with the magnetic 
field i.e. it
is exponentially small in the field and becomes significant when  
$\sqrt{v_x/v_y}l_B\sin{k_0a} \sim 1$. 
Depending on the anisotropy in the Weyl cone dispersion, the behavior of 
the induced gap alters significantly. There are two distinct regimes 
which are distinguished by the anisotropy parameter $\gamma$, defined in Eq. 
\ref{Eq:Anisotropy_pm}.  Here we summarize the results in the case $\phi_B\ll \phi_0$ i.e. the flux through a 2D unit cell of area $a^2$ is much smaller than
the flux quantum $\phi_0=h/e$, which is the relevant regime in typical experiments.  In later sections, we will also discuss the very large field regime ($\phi_B\sim\phi_0$).

(i) For $\gamma<0$, the system displays three phases as a function of the separation between the Weyl nodes, $Q$, see Fig.~\ref{Fig:Phase_Schematic}(a3).  For small $Q$, the system is a normal insulator (NI), whereas for large $Q$ it is a layered Chern insulator (LCI). The LCI phase 
can be viewed as a collection of two-dimensional Chern insulators stacked along the $x$ direction, hosting
chiral surface states on surfaces that result from termination along the $y$ and $z$ directions.

We denote the critical momentum space separation at which the transition occurs by $Q^*$. 
These two insulating phases are separated by a gapless WSM phase.
The width (range of $Q$ values) of the WSM region is exponentially small in $\phi_B^{-1}$. This is associated with the finite bandwidth 
of the LLs which is exponentially small in $\phi_B^{-1}$.  For fixed $Q$ inside the insulating phases, the gap increases monotonically with the field.

(ii) For $\gamma>0$, as $Q$ is increased from zero, the system alternates between NI and a distinct insulating phase, which we denote LCI$'$, and which can be viewed as two copies of LCI with opposite Chern numbers.  LCI$'$ is a symmetry-protected topological phase relying on translational and mirror symmetries -- in their absence, the phase is indistinguishable from NI.  This alternation continues until $Q$ reaches a value $Q^*_{\mathrm{osc}}$, beyond which the system enters an LCI state, see Fig.~\ref{Fig:Phase_Schematic}(b3).  The transitions between NI and LCI$'$ phases are separated by finite-size regions (that are exponentially narrow in $\phi_B^{-1}$) consisting of three phases: two WSM states 
which are separated by a gapped LCI state.  
For fixed $Q<Q^*_{\mathrm{osc}}$, the induced gap oscillates with the applied field and vanishes in an almost periodic manner (strictly periodic in the continuum). The envelope of oscillations decays 
exponentially with $(Ql_B)^2$.

It is worth emphasizing which of these effects rely on the presence of a lattice.   First of all, the appearance of an LCI phase at large $Q$ (for either case) is directly attributed to the periodic nature of the BZ.   In addition, the appearance of WSM (for $\gamma<0$) and WSM-LCI-WSM regions (for $\gamma>0$) separating the insulating phases only occurs on a lattice, where the LLs develop a dispersion in momenta perpendicular to the direction of the field.  Finally, the identification of the topological nature of the LCI and LCI$'$ phases is only possible in a lattice model.  

In addition, the fact that we work with a lattice model enables us to make a systematic
study of the fate of the Fermi-arc states in the presence of commensurate magnetic 
fields. For the case $\gamma<0$, we find that the Fermi-arc states are either 
immediately gapped out or evolve to a Fermi loop which extends to the 
edges of the surface BZ. For $\gamma>0$, the situation is more involved 
because of the oscillations and periodic appearance of the WSM state. 
However in either case, the fate of Fermi-arc surface states can be predicted from the topology of the bulk states, as will be discussed below.

\section{Continuum description}
\label{Sec:Continuum}

In this Section, we present a unified description of the low-energy continuum 
behavior of our model, previously discussed in Refs. \cite{Chan_Lee_2017, 
Saykin_Rodinov_2018, Devakul_Parameswaran_2021}.

A single Weyl fermion centered at momentum ${\bf Q}/2$ is described by the Weyl Hamiltonian, $H({\bf k}) = \sum_i v_i (k_i-Q_i/2) \sigma_i$, where ${\bf v}=(v_x, v_y, v_z)$ is the velocity. In the presence of a quantizing orbital 
magnetic field, ${\bf B} \parallel \hat{z}$, the spectrum consists of graphene-like LLs $E_n = 
\pm \sqrt{\frac{2|v_x v_y|}{l_B^2} n + v^2_zk_z^2}$ ($n=1, 2, ...$).  In addition, the spectrum
contains a special chiral LL which disperses linearly in momentum 
along the direction of the field, $E_0 = \eta v_z k_z $, with $\eta=\mathrm{sgn}(v_x v_y v_z)$ the chirality of the Weyl fermion. The gapless chiral LL is responsible for phenomena associated with chiral anomaly physics, including  negative magneto-resistance, thickness-dependent quantum oscillations, and 3D quantum Hall effect in Weyl semimetals.

In  WSMs, Weyl fermions always come in pairs of opposite chriality,  separated 
by a finite distance $Q$ in momentum space, and by a finite energy barrier. 
This gives a correction to the low energy spectrum due to the tunneling of 
electrons from one WN to the other. This correction is exponentially small 
in $Q\, l_B$ and hence nonperturbative in field. The effect will be significant 
when the cyclotron gap  $\sqrt{2|v_x v_y|}/l_B$  is comparable  to the height 
of the barrier, which can be approximated by $ v_x Q/2 $ assuming that the 
WNs are separated in the $k_x$ direction. Therefore the condition for states 
around the two WNs to be strongly coupled can be expressed as 
\begin{align}\label{Eq:Condition_Coupling}
 \sqrt{\abs{\frac{v_x}{v_y}}} Q l_B \lesssim 1 , 
\end{align}
which depends on the anisotropy of the Weyl cone. The above 
condition was derived before in Ref. \onlinecite{Chan_Lee_2017}. 

The condition 
for chiral LLs of two Weyl nodes to be strongly coupled can also be stated 
in terms of the overlap of their wavefunction in momentum space. 
In the Landau gauge, ${\bf A}=(-y, 0, 0)B$,  the wavefunction of the chiral LL associated with a Weyl node at $\pm{\bf Q}/2$ is $|ZLL\rangle \sim \exp(- \abs{\frac{v_x}{v_y}} \frac{l_B^2}{2}(k_x + y/l_B^2\mp Q/2)^2)$. 
The overlap of these two wavefunctions is $\propto\exp(- \abs{\frac{v_x}{v_y}} \frac{l_B^2}{4}Q^2)$
and becomes significant when $\sqrt{\abs{v_x/v_y}} 
Ql_B \lesssim 1$, which is identical to the condition in 
Eq. \ref{Eq:Condition_Coupling}.

The preceding discussion gives only a qualitative picture of coupling 
of two WNs and the potential gap opening in the spectrum in presence 
of an orbital magnetic field directed perpendicular to their separation. 
Below we consider a low-energy model of WSM and discuss 
quantitatively how the field-induced energy gap behaves with the magnetic field and the anisotropy parameter $\gamma$. 

\begin{figure}
\includegraphics[width=1\linewidth]{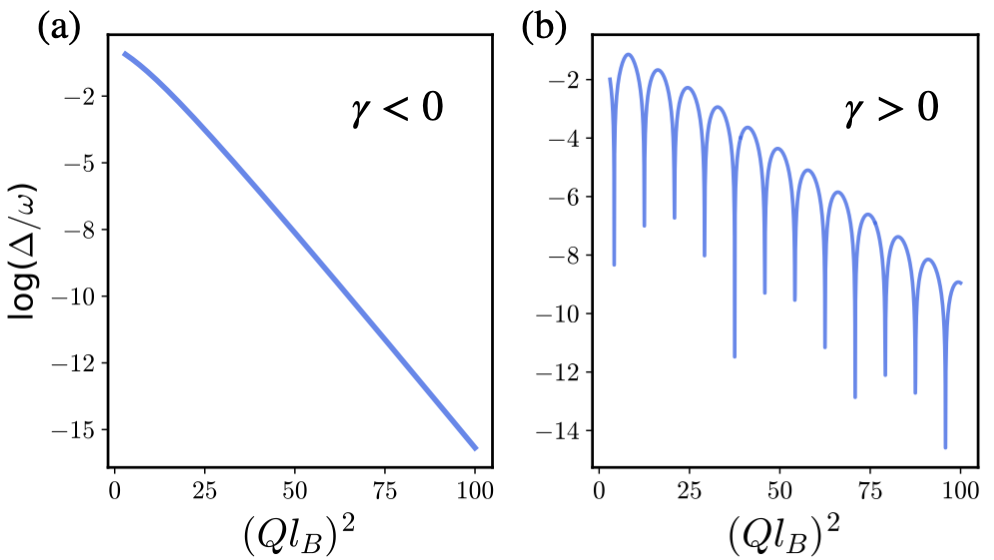}
\caption{
Energy gap ($\Delta$) in the continuum model as a function of the 
square of the dimensionless parameter $Ql_B$, computed numerically for $t=1$. 
(a) For $\gamma=-0.44$, the field-induced gap falls exponentially with $Ql_B$.  
(b) For $\gamma=0.96$, the gap oscillates and periodically 
goes to zero (at the troughs). The troughs in the figure are not exactly 
at zero because of the finite line grid in the numerical computation. 
Here the envelope of oscillations falls exponentially.}
\label{Fig:GapContinuum}
\end{figure}

\subsection{Low energy continuum model}

When the separation $Q=2k_0$ between the Weyl cones is small compared to 
the linear dimension of the BZ, the lattice Hamiltonian (Eq. \ref{Eq:LatWsm}) 
can be approximated to 
\begin{align}\label{Eq:HW_Continuum}
     H({\bf k}) = \tilde{v}_x m({\bf k}) \sigma_x
     + 2 v_y k_y \sigma_y + 2 v_zk_z \sigma_z,
\end{align}
where now $m({\bf k})=(k_x^2-k_0^2) + t^2(k_y^2 +k_z^2)$, and $\tilde{v}_x=v_x/k_0$.   
For a field along the $z$ direction, momentum along $z$ remains 
conserved. We are interested in the energy gap at the band bottom $k_z=0$.

As we have already mentioned, the anisotropy  parameter 
$\gamma$ (defined in Eq. \ref{Eq:Anisotropy_pm}) controls the structure 
of the gap as a function of $Ql_B$. For an arbitrary $t$, the problem of 
computing the gap analytically becomes challenging because one would require 
solving a fourth-order differential equation. 
However, as shown in Ref. \onlinecite{Devakul_Parameswaran_2021}, a wealth of 
information on how the induced gap behaves as a function of magnetic flux 
or WN separation can be extracted analytically without solving the full 
eigenvalue problem.

\subsection{Zero energy solutions}
\label{sec:zeroEnergy}

Following Ref.~\onlinecite{Devakul_Parameswaran_2021} it is possible to obtain analytic conditions for the system to remain gapless, even in the presence of a magnetic field.  The authors in Ref.~\onlinecite{Devakul_Parameswaran_2021} considered the case of $t=1$.
Here we generalize the treatment to arbitrary $t\ne 0$ (the case $t=0$ is discussed in detail below).

For an orbital magnetic field along $z$ direction and in the gauge 
${\bf A}=(-y, 0, 0)B$, the momenta $k_x$ and $k_z$ remain good 
quantum numbers. External magnetic field is introduced via minimal coupling
$k_x \to (k_x + y/l_B^2)$.
Introducing the ladder operators $a=\frac{l_B}{\sqrt{2t}}(k + 
\frac{t}{l_B^2}\frac{d}{dk} )$, $a^{\dagger}=\frac{l_B}{\sqrt{2t}}(k - 
\frac{t}{l_B^2}\frac{d}{dk} )$, where $k = (k_x+ y/l_B^2)$, 
the Hamiltonian in the presence of a magnetic field can be expressed as 
\begin{align} \label{Eq:HBCon}
    H_{\phi} = \omega \begin{pmatrix}
        0 & h^{\dagger} \\
        h & 0 
    \end{pmatrix}
\end{align}
where $h = \left[a^{\dagger}a - \delta + \eta(a-a^{\dagger}) \right]$
,  $\delta=\frac{k_0^2l_B^2}{2t} - \frac{1}{2}$,  $\eta=\frac{k_0l_B}{\sqrt{2t^3}} \frac{v_y}{v_x}$, and 
$\omega=\frac{2v_x t}{k_0l_B^2}$.  Note that $H_\phi^2=\omega^2\left(\begin{array}{c c}
h^\dagger h& 0 \\ 0 & h h^\dagger\end{array}\right)$.  Hence, $H_\phi$ has a zero-energy 
eigenvalue if and only if $h$ has a zero eigenvalue. The eigenvalues  of  $h$ are 
\begin{equation}
    \lambda_N=
N+\frac{1}{2}-\frac{k_0^2l_B^2}{2t} \gamma,
\end{equation}
where $N=0, 1,2,...$,  and $\gamma=\big(1-v_y^2/(v_x^2t^2)\big)$ as defined 
in Eq.~\eqref{Eq:Anisotropy_pm}.  

Clearly, the spectrum does not contain
zero-energy eigenvalues when $\gamma$ is negative.  In this case, the 
spectrum is always gapped for any finite values of magnetic field.  
On the other hand, for positive $\gamma$, zero-energy eigenvalues 
occur periodically in  $k_0^2l_B^2$ at the values $\gamma k_0^2l_B^2=t(2N+1)$.
Therefore, the gap closes periodically with $(k_0l_B)^2$. 
The period of oscillation in $l_B^2$ is $T_{l_B^2} =\frac{2\pi}{\gamma A}$, where 
\begin{equation}\label{eq:A}
 A=\pi k_0^2/t.   
\end{equation}
For $v_y=0$, the model in Eq. \ref{Eq:HW_Continuum} describes a nodal 
loop semimetal and then $A$ is the area of the nodal loop $k_x^2 + t^2k_y^2=k_0^2$.

\subsection{Structure of the induced gap}

To find the behavior of the energy gap $\Delta$ as a function of field and WN separation $Q$, we numerically diagonalize the 
Hamiltonian in Eq. \ref{Eq:HBCon} in a LL basis by truncating to a 
hundred Landau levels. We notice that the separation parameter $k_0=Q/2$ 
and magnetic length $l_B$ always appear as a dimensionless product 
$k_0l_B$ in the block $h$. Therefore the spectrum and energy gap of 
$H_{\phi}/\omega$ are functions of the dimensionless parameter $Ql_B$.   The numerically computed 
energy gap $\Delta$ is plotted in Fig. \ref{Fig:GapContinuum}. 
We find that $\Delta/\omega$ decays exponentially in $(Ql_B)^2$.  For $\gamma<0$, this decay is monotonic and the gap never closes. By contrast, for $\gamma>0$, 
the gap vanishes periodically with $(Ql_B)^2$, in agreement with the results of Sec.~\ref{sec:zeroEnergy}.

We can understand the monotonic regime from the special case $t=0$, which was studied in Ref. \cite{Saykin_Rodinov_2018}. 
Note that for $t=0$, the anisotropy parameter $\gamma$ is 
always negative for any finite value of the ratio  $v_y/v_x$. 
For $t=0$, the problem of finding the spectrum 
in a uniform magnetic field along $z$ can be mapped to a
Schrodinger equation with an asymmetric double-well potential. The 
authors in Ref. \cite{Saykin_Rodinov_2018} found that the induced gap is 
nonperturbative in magnetic field. They computed the energy gap ($\Delta$) 
using the WKB  approximation and found it to be exponentially small in 
$Q l_B$:
\begin{align}\label{eq:gap_cont}
    \Delta \approx \frac{v_x}{\sqrt{\pi}} \sqrt{\frac{v_x}{v_y}}
    \frac{1}{l_B} e^{-\frac{1}{6} \frac{v_x}{v_y} (Ql_B)^2}.
\end{align}
The orbital magnetic field immediately introduces a finite gap 
in the electronic spectrum (i.e. gapping the WNs) which increases 
monotonically with  $B \propto 1/l_B^2$. We note that the gap $\Delta$ 
is significant only when the condition  \eqref{Eq:Condition_Coupling} 
is satisfied.

\subsection{Discussion}
\label{Subsec:DisCon}

Several comments are in order. The oscillation and the periodic closing of 
the energy gap cannot be understood from the Onsager theory; its origin is 
purely a quantum phenomenon. The Fermi surface at zero energy consists 
only of two isolated points- the two Weyl nodes. At energies away from 
the energy of WNs, the Fermi surface consists of two disjoint and 
identical Fermi pockets around the Weyl nodes in the $k_x$-$k_y$ plane. 
When an orbital magnetic field is applied it induces a quantum tunneling from one Fermi pocket 
to the other, a phenomenon known as magnetic breakdown \cite{Cohen_1961, Kaganov_1983, Shoenberg_1984}.
It is this quantum tunneling between the two Fermi pockets around the WNs which leads to a gap opening in the spectrum. 
The probability of 
tunneling is exponentially small in the distance between the Fermi pockets and the height of the barrier, giving rise to the exponential decay in $(Ql_B)^2$.
For $\gamma>0$ the Fermi pockets are in crescent shape 
with two close encounters as shown in Fig. \ref{Fig:Phase_Schematic}(b2). This gives two possible tunneling paths for the magnetic breakdown. The interference between these gives rise to the periodic gap closings. The area enclosed by the two paths is $A$ as defined in Eq.~\eqref{eq:A} accounting for the period of oscillations. 
Note that this picture relies on mirror symmetry which ensures that the two tunneling paths have exactly the same amplitude. If the symmetry were absent, the destructive interference would be imperfect, and the periodic gap closings would not occur.
On the other hand for $\gamma<0$ the Fermi pockets have an 
elliptical shape as shown in Fig. \ref{Fig:Phase_Schematic}(a2) with only one close encounter between them, leading to a monotonic behavior of the gap.

The continuum calculation provides a good description only when the separation $Q$ 
between the WNs is much smaller than the linear dimension of the BZ i.e. $Q \ll \pi/a$ 
and for $l_B\gg a$. We devote following sections to study the effects of the lattice.

\section{Topological nature of the gapped phases on a lattice}
\label{Sec:Lattice}

The lattice model allows us to discuss the topological nature of the 
gapped phases induced by orbital magnetic fields.  It is instructive 
to start from the Weyl semimetal in 
the absence of a magnetic field.  Then, one can introduce the $k_x$-dependent 
Chern number $C(k_x)$, which measures the Chern number in the $k_y$-$k_z$ 
plane at fixed $k_x$ of the occupied bands.  For the lattice model defined 
in Eq. \ref{Eq:LatWsm}, 
$\sum_{n \in occ} C_n(k_x)=1$ for all values of $k_x$ in between the two Weyl nodes, 
$-k_0<k_x<k_0$, and it is zero otherwise.  For a system in a slab 
geometry with a surface in the $xy$ plane, this gives rise to Fermi 
arc states on the surface connecting the projection of the two Weyl 
nodes on the $k_x$-$k_y$ surface BZ.

\begin{figure}
\includegraphics[width=1\linewidth]{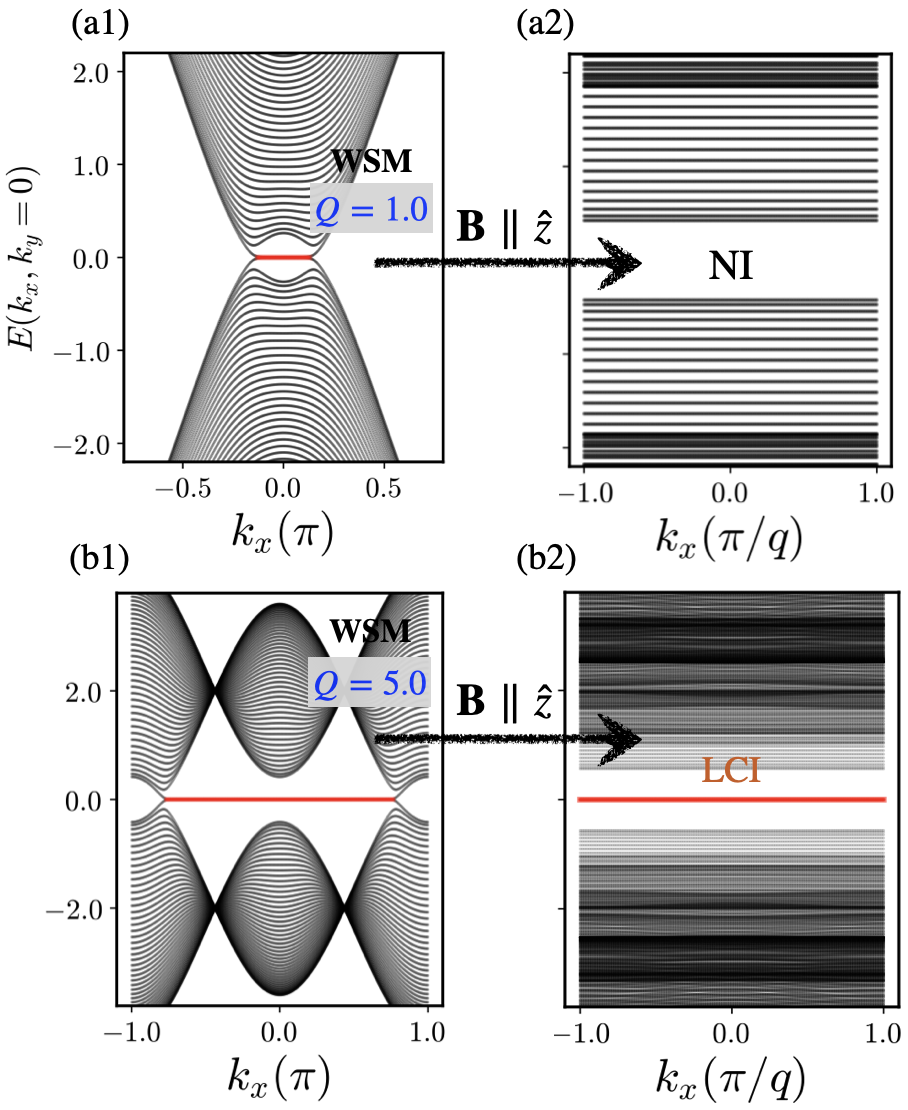}
\caption{Energy spectrum (along $k_x$) in a slab geometry along $z$ 
direction with thickness $L_z=50$ in units of lattice constant for 
$\gamma \approx -0.30$ (negative). Energies are given in units of $v_x$. 
Left column shows the slab spectrum for zero field, highlighting 
the zero energy Fermi-arc surface states (a1) for small  ($Q=1$)
and (b1) large ($Q=5$) separation between the Weyl nodes. 
Right column shows the slab spectrum in presence of an external 
field applied along $z$ direction corresponding to a flux of 
$\phi_B/\phi_0=1/q$ with $q=10$. (a2) Fermi-arc states get fully 
gapped out (NI phase). (b2) The Fermi-arc states of the
WSM evolve to the surface states of a LCI phase (highlighted 
in red). The surface states in the LCI phase are extended to the edge of 
the magnetic BZ along $k_x$. }
\label{Fig:Surface_Mono}
\end{figure}

We now consider the effect of a magnetic field. When the two WNs are 
close to the center of the BZ, the Fermi arcs are short 
(see Fig.~\ref{Fig:Surface_Mono}(a1)). We then intuitively expect the 
WNs to gap out due to $Q$ tunneling (tunneling within the first BZ), 
and the Fermi arcs to disappear in the presence of a magnetic 
field \cite{Abdulla_2024}.  This corresponds to a topologically trivial 
insulating phase, which we refer to as normal insulator (NI).

On the other hand, when the Weyl nodes approach each other across the edges 
of the BZ, the Fermi arcs are long, nearly spanning the full BZ in the $k_x$ 
direction (see Fig.~\ref{Fig:Surface_Mono}(b1)). Now, when we apply a magnetic 
field, we expect the Weyl nodes to gap out due to $Q'$ tunneling (tunneling 
across the edge of the BZ), but the gapless Fermi arc states to 
survive \cite{Abdulla_2024}. This corresponds to the LCI phase. 

\begin{figure}
\includegraphics[width=1\linewidth]{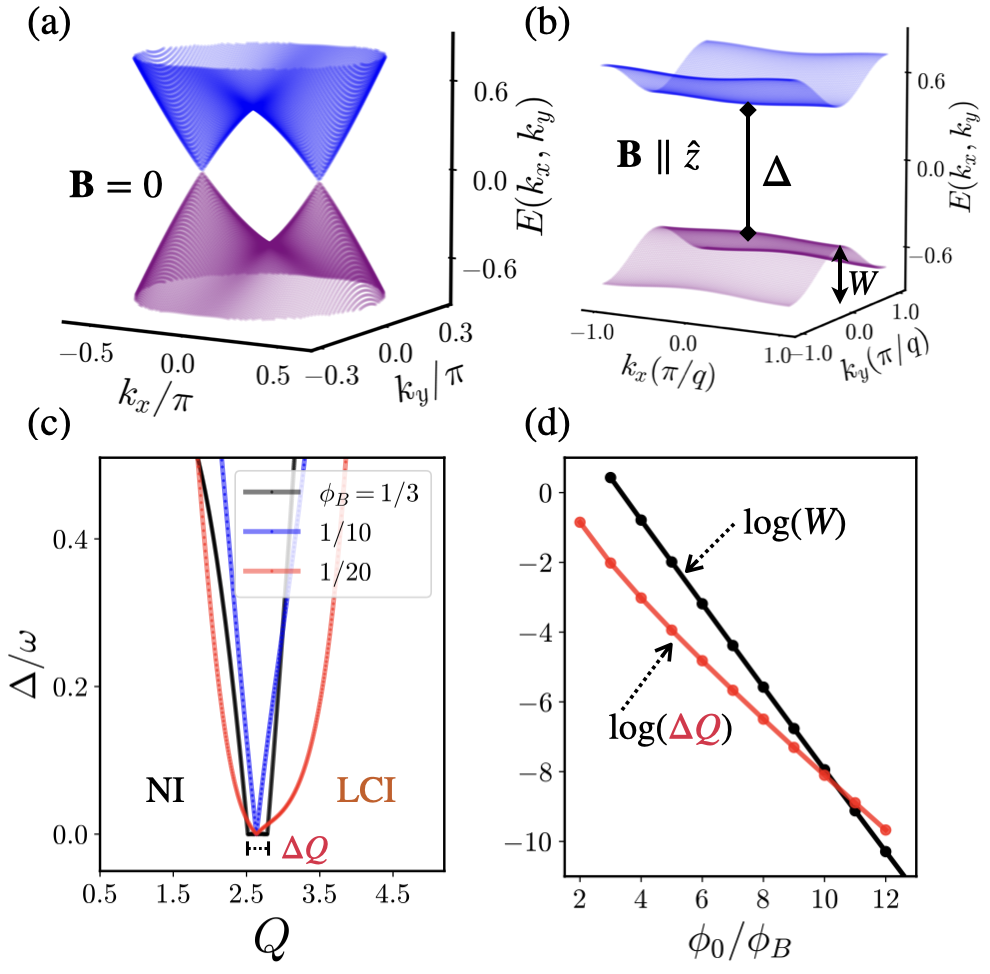}
\caption{ Behavior of the gap for $\gamma<0$. (a) Zero field energy spectrum 
is showing two Weyl cones separated along $k_x$. (b) In presence of an 
orbital ${\bf B}\parallel \hat{z}$,  Weyl cones 
are gapped out. Only two Landau bands near zero energy are shown. For 
flux $\phi_B \sim \phi_0$, the bands have a finite bandwidth ($W$). 
(c) Energy gap ($\Delta$) as a function of Weyl nodes' separation for flux 
values $\phi_B/\phi_0 = 1/q=1/3, 1/10, 1/20$.   
Magnetic field immediately creates a gap in the spectrum for almost any 
value of separation between the Weyl nodes. The gap is exponentially 
small $\sim \exp(-Q^2l_B^2)$. For intermediate values of separation, 
WSM state survives  but in an exponentially small vicinity of $k_0$ 
(seen most clearly for $\phi_B/\phi_0=1/3$). (d) The width
of the gapless region (denoted by $\Delta Q$ in (c))  where the system 
remains in the WSM phase falls exponentially with $\phi_B^{-1}$. 
The bandwidth (denoted as $W$ in (b)) also falls 
as $\exp(-\phi_B^{-1})$.}
\label{Fig:Gap_Phase_Lat_Mon}
\end{figure}

We note that this behavior is expected independently of the sign of $\gamma$.
These notions can be made more concrete by introducing the magnetic BZ, which we discuss next.  In the magnetic BZ, the momenta are good quantum numbers, making it possible to track the evolution of the Weyl nodes and Fermi arcs in the presence of a magnetic field.  Figure \ref{Fig:Surface_Mono}(a2,b2) shows the spectrum on the magnetic BZ for the $Q$ and $Q'$ tunneling cases, respectively, where the NI and LCI phases are observed.  These results will be discussed in more detail below. Interestingly, the more careful analysis will also reveal the appearance of an additional gapped insulating phase, LCI$'$, which alternates with the NI phase in the case $\gamma>0$.

As shown in this section, the behavior of the Fermi-arc surface states 
discussed above in the presence of a magnetic field is consistent with the 
nonperturbative gap-opening mechanism in the bulk Weyl semimetal. Fermi-arc 
states with a short arc length disappear once the magnetic field is applied. 
In this regime, the surface states acquire a finite lifetime, as electrons 
initially localized at the surface can escape into the bulk through chiral 
Landau levels. This fate of short Fermi arcs has also been predicted in 
a recent continuum analysis \cite{Francesco_2014}. In contrast, Fermi-arc 
states with a long arc length persist and smoothly evolve into the chiral 
surface states of the LCI phase.

\subsection{Bloch-Hofstadter Hamiltonian}

In what follows we will set $t=1$ in our lattice model.  
In presence of an orbital magnetic field along $z$ direction 
(i.e. perpendicular to the direction of separation $\vec{Q}$), 
and using the Landau gauge ${\bf A} = (-y, 0, 0)B$, 
the Hamiltonian in Eq. \ref{Eq:LatWsm} takes the 
following form
\begin{equation}
\begin{aligned} \label{Eq:LatB}
    H_{\phi} = \sum_y \sum_{k_x, k_z} &  c^{\dagger}_y ~ 2\left[f_1(y)\sigma_x 
    + f_2(y) \sigma_z \right]c_y \\ 
    & - \left(c^{\dagger}_{y+1}\left(\tilde{v}_x\sigma_x +iv_y \sigma_y \right)c_y 
     + H.c \right).
\end{aligned}
\end{equation}
Here $f_1(y) = 
\tilde{v}_x\left(2+\cos(k_0a) - \cos(k_za) - \cos(k_xa + 2\pi\phi_B y)\right)$, $f_2(y) = f_2 = v_z \sin(k_za)$, and $\phi_B = Ba^2/\phi_0$ ($\phi_0=h/e$) 
is the magnetic flux through a 2D unit cell in the $x$-$y$ plane.  
Unless otherwise stated, hereafter we set $v_x=1$ and vary $v_y$ to control $\gamma$. 

Translation symmetry can be restored when the flux is commensurate with the lattice.  
This is done by choosing an enlarged unit cell containing an integer 
number of flux quanta. In what follows we will 
choose commensurate flux $\phi_B/\phi_0=p/q$, 
where $q$ is a positive integer and without loss of generality we will set $p=1$ 
throughout this paper. The enlarged unit cell is then obtained by extending the 
original unit cell $q$ times along the $y$ direction. This gives a magnetic 
BZ with $-\pi \le k_xa \le \pi$, $-\pi/q \le k_ya \le \pi/q$, $-\pi \le k_z a 
\le \pi$. 
Working with the magnetic BZ will allow us to compute topological invariants 
and track the evolution of the Fermi arc states in a straightforward manner.
Although $k_x$ in the magnetic BZ lies in the range $(-\pi,\pi)$ the spectrum 
along $k_x$ is in fact periodic with a period of $2\pi/q$ \cite{Bernevig2013}. 
Therefore, in the analysis below we restrict ourselves to $k_x\in (-\pi/q,\pi/q)$. 

Here we emphasize that, in contrast to Landau levels in the continuum, 
the Hofstadter Bloch bands arising at commensurate magnetic flux can host Weyl 
nodes and carry nonzero Chern numbers. Throughout this work, when we refer 
to a WSM phase in the presence of a magnetic field, we specifically 
mean that the corresponding Hofstadter Bloch bands exhibit Weyl nodes 
pinned at zero energy.  \\

In all plots of the energy gap in this section we plot it in units of $\omega$ 
defined previously for the continuum model and explicitly given by 
$\omega=2\pi v_x/(q k_0) $ for the lattice model.

\begin{figure}
\includegraphics[width=1\linewidth]{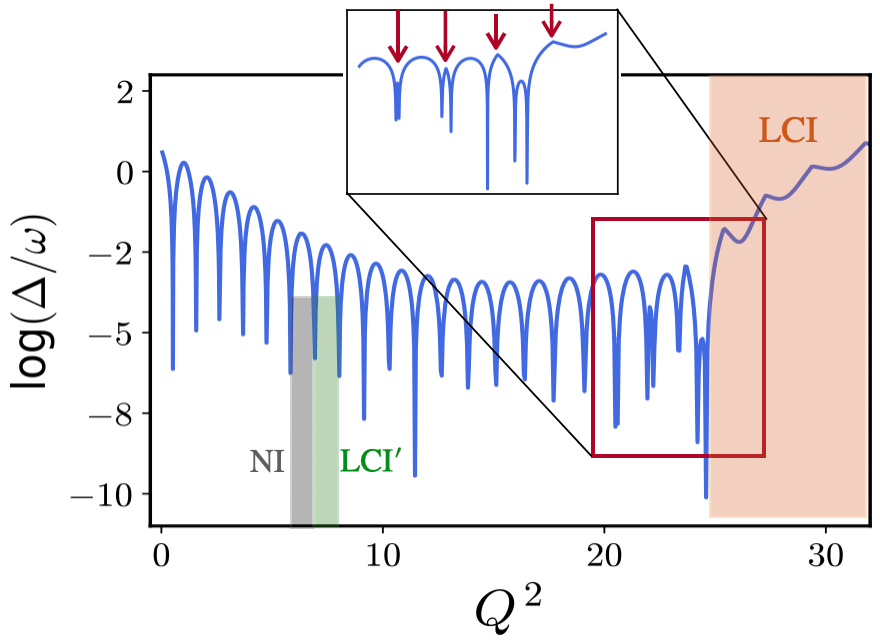}
\caption{Logarithm of the energy gap $\Delta$ as a function of the square
of the separation between the Weyl nodes for a fixed flux 
$\phi_B/\phi_0=1/50$, and for $\gamma=0.96$ (positive). The troughs are not exactly at 
zero because of a numerical artifact: finite sampling for $Q$ as well as 
for momenta. Periodic closing of gap stops after a certain value of 
$Q^2 \approx 24$. Near that value, besides the smooth peaks, we observe 
additional sharp peaks (shown in the inset). For $Q \gtrsim 24$, the system is 
in an LCI state. For $Q \lesssim 24$, the WSM under magnetic field alternates 
between an NI and an LCI$'$ which can be viewed as two copies of an LCI with 
opposite Chern numbers. }
\label{Fig:Log_gap_q50}
\end{figure}

\subsection{Phase diagram for $\gamma<0$}
\label{Subsec:ZES}

For the special case $\tilde{v}_x=v_y=1$, 
we can understand the transition between NI and LCI analytically
by solving for the zeros of the Bloch-Hofstadter Hamiltonian in Eq. \ref{Eq:LatB}.
These correspond to Weyl nodes 
in the magnetic BZ \cite{Abdulla_Murthy_2022, Abdulla_2024}, and their 
appearance is determined by the condition
\begin{align}\label{Eq:WeylSol}
    \cos{qk_x} = T_q(g) - 2^{q-1},
\end{align}
where $T_q(g)$ is Chebyshev polynomial of the first kind of degree $q=\phi_0/\phi_B$, 
and $g=1+\cos{k_0}$. Since $0 \le k_0 \le \pi$, the parameter $g$ lies in the 
range $0 \le g \le 2$.

\begin{figure*}
\includegraphics[width=1\linewidth]{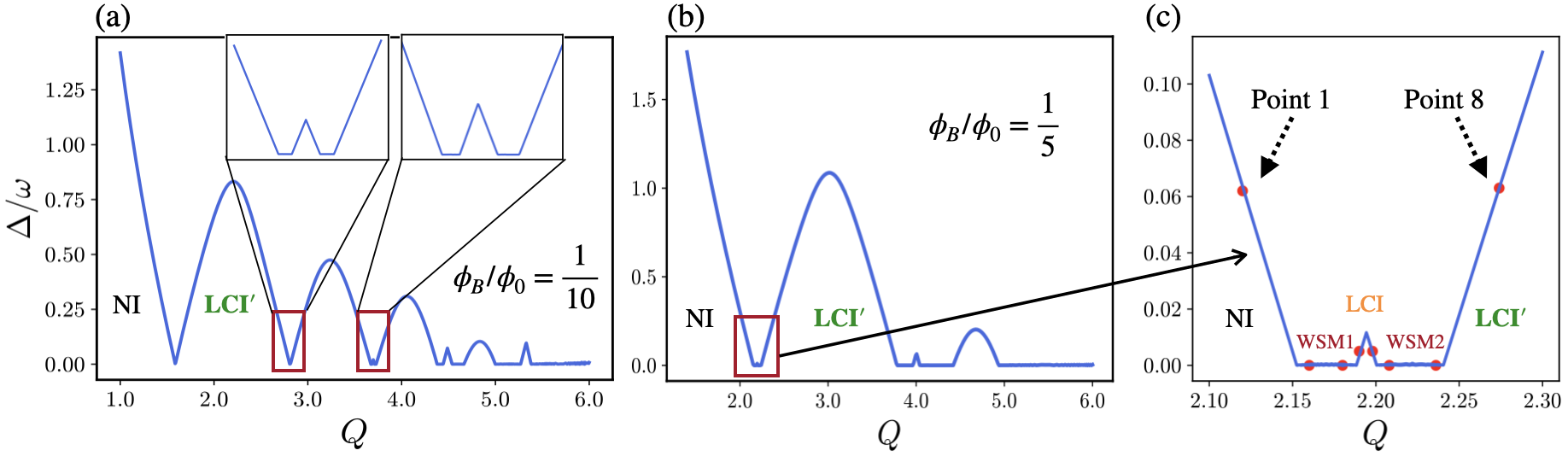}
\caption{Energy gap as a function of separation between the Weyl nodes 
for sizable fluxes (a) $\phi_B/\phi_0=1/10$ and (b) $\phi_B/\phi_0=1/5$ for 
a fixed $\gamma=0.96$. See text for a detailed discussion of the sequence 
and structure of the transitions between the different gapped phases. (c) In 
a transition from an NI to an LCI$'$ the system goes through several 
closings and reopenings of the gap, as shown in Fig. \ref{Fig:Band_Evolution}. 
The red dots in (c) correspond to the values of $Q$ at which the energy bands in Fig.~\ref{Fig:Band_Evolution} are plotted.}
\label{Fig:Gap_q5_10}
\end{figure*}

\begin{figure*}
\includegraphics[width=0.9\linewidth]{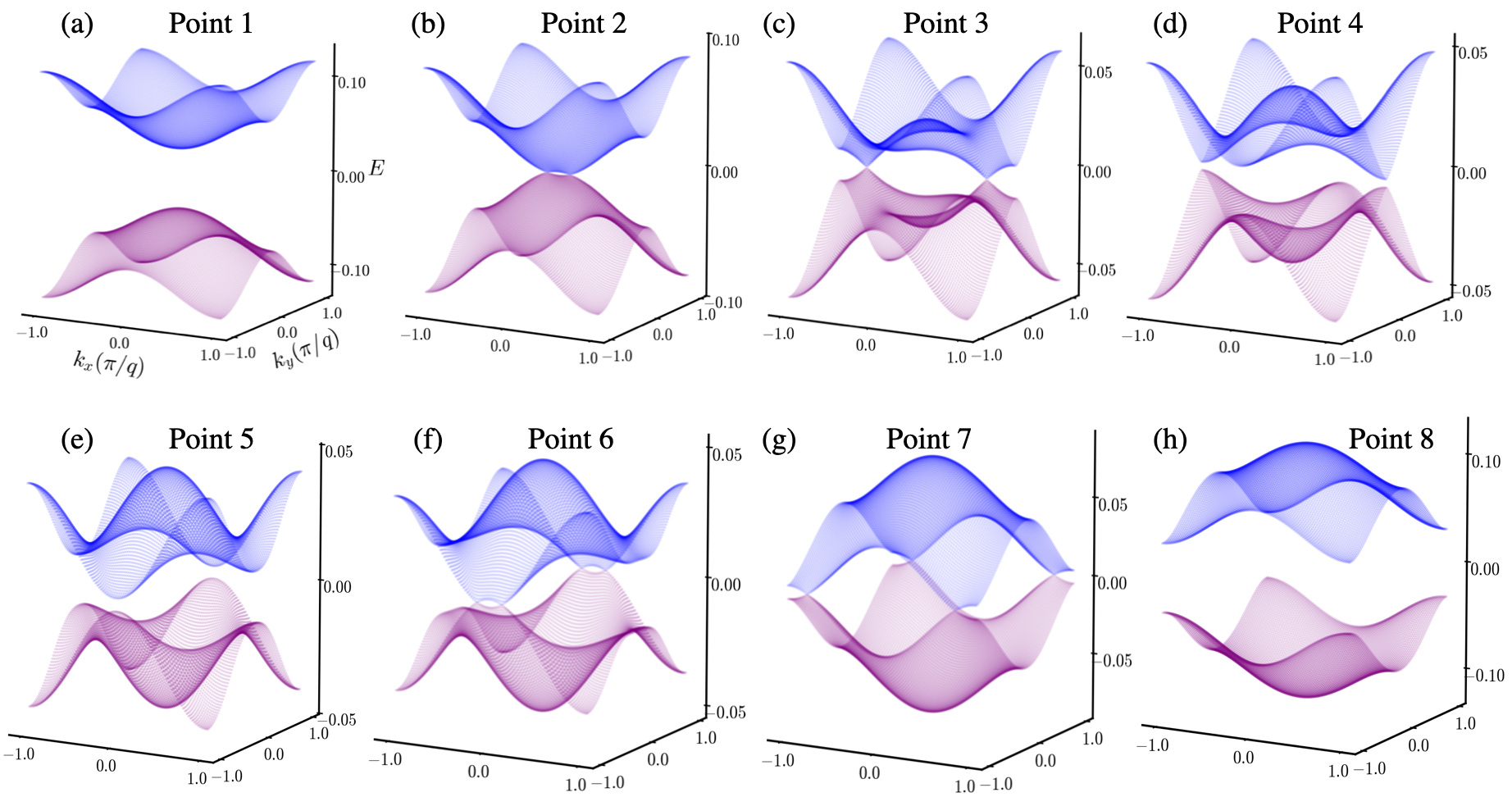}
\caption{Gap closing, formation of Weyl cones, and gap reopening across the transition 
between the NI and LCI' states, as $Q$ is increased, shown in Fig.~\ref{Fig:Gap_q5_10}(c). Here the flux is fixed and equal to $\phi_B/\phi_0=1/5$. 
The values of $Q$ at which the energy bands are plotted are indicated by red dots in Fig.~\ref{Fig:Gap_q5_10}(c). The transition from (a) to (d) involves 
a band inversion at $(0,0)$, and the transition from (e) to (h) involves 
another band inversion at $(\pi/q, \pi/q)$. }
\label{Fig:Band_Evolution}
\end{figure*}

Equation \eqref{Eq:WeylSol} determines when the WSM under magnetic field 
becomes gapped or remains gapless.  For certain values of $q$ and $g$, 
the equation has no 
solutions, which indicates a gapped phase.   On the other hand, if the 
right-hand-side of the equation lies in the range $(-1, 1)$, then there are 
real solutions, indicating that the spectrum contains two WNs at $(\pm k_1, 0, 0)$, 
where $0 < k_1 < \pi/q$ is the solution for $k_x$ in Eq.~\eqref{Eq:WeylSol}. 

Consider first the limit of large $q$, corresponding to a small flux.  
In this limit, $2^{q-1}$ is exponentially large and we can neglect the 
cosine term by comparison.  Then,
\begin{equation}
\begin{aligned}
    T_q(g) \approx 2^{q-1} .
\end{aligned}
\end{equation}
Since $-1<T_q(g)<1$ for $g\le 1$, the above equation can admit a solution 
only when $g>1$. For $g>1$, we have  $2T_q(g) = 2\cosh(q \cosh^{-1}(g)) = 
\exp(q\ln(g+\sqrt{g^2-1}))+ \exp(-q\ln(g+\sqrt{g^2-1}))$. Substituting 
this expression of $T_q(g)$ we convert the above condition into a quadratic
equation. 
This equation admits a single solution, $g=5/4$. Thus, in the limit of large $q$, the system is gapped for all values of $Q=2k_0$ except for $Q^{*}=2k_0^* = 2\cos^{-1}(1/4)$.

Reintroducing the cosine results in a finite range of $g$ where the system is 
gapless, $g_1<g<g_2$, where $g_1$ and $g_2$ are critical values surrounding 
$g=5/4$.  This corresponds to the WSM phase that separates the two insulating 
states. The width of this region $g_2-g_1$ is of order $2^{-q}$, i.e. it is
exponentially small in $1/\phi_B$, as shown in Fig. \ref{Fig:Gap_Phase_Lat_Mon}(e).

To understand the topological nature of the gapped phases, we calculate the $k_x$-dependent Chern number $C(k_x)$ in the magnetic BZ. We find that for large $g$ (i.e. small $Q<Q^*$) $\sum_{n \in occ} C_n (k_x)=0$ for all $k_x$, corresponding to an NI. On the other hand, for small $g$ (i.e. large $Q>Q^*$), $\sum_{n \in occ}C_n(k_x)=1$ for all $k_x$, corresponding 
to an LCI phase.
The transition between NI and LCI can be understood from the evolution of the solutions to Eq. \eqref{Eq:WeylSol}. Let us fix the flux $1/q$ and look for solutions as 
$g$ (or equivalently $Q$) is varied.  Starting from large $g$ (NI phase), when $g$ is decreased to $g_2$, the band gap closes and a Dirac point appears at the Gamma point of the magnetic BZ. Decreasing $g$ further leads to a splitting of the Dirac point into a pair of WNs that move away from each other towards the zone boundary. In this regime, solving in the slab geometry, a Fermi arc appears at the surface, and its length increases with the separation between the WNs. At a second critical value, $g=g_1$, the two WNs merge at the edge of the magnetic BZ $(\pm \pi/q, 0, 0)$ and disappear but the Fermi arc survives and spans the full magnetic BZ along $k_x$. Thus an LCI state emerges. This is demonstrated in Fig.~\ref{Fig:Surface_Mono}.

Even though the exact solution above was derived for $\tilde{v}_x=v_y=1$, 
we confirm that this behavior is generic for all $\gamma<0$ by numerically 
solving the Bloch-Hofstadter Hamiltonian, Eq.~\eqref{Eq:LatB}. This is 
demonstrated in Fig.~\ref{Fig:Gap_Phase_Lat_Mon}. We note that the finite 
width of the WSM phase can be attributed to the finite bandwidth of the 
Landau levels on a lattice.  The width of the gapless region and the bandwidth 
are plotted in Fig.~\ref{Fig:Gap_Phase_Lat_Mon}(e,f). As can be seen, they 
both show an exponential dependence $\sim \exp(-\phi^{-1}_B)$.

\subsection{Phase diagram for $\gamma>0$ and appearance of LCI$'$}
\label{Subsec:LargeFlux}


We now turn to $\gamma>0$. Recall that in the continuum picture, closings and reopenings of the gap occur in an oscillatory fashion as either $Q$ or the flux $\phi_B$ is varied. As discussed in Sec.~\ref{Subsec:DisCon}, this can be understood from the semiclassical picture of destructive interference between two (intra-BZ) tunneling paths. On the lattice, there is an additional tunneling path available, inter-BZ tunneling.   As the WNs approach the BZ edge, inter-BZ tunneling becomes dominant. Therefore, for large $Q$ we expect the oscillations in the gap to stop and the system to enter an LCI phase. Furthermore, the inter-BZ tunneling affects the structure of the gap even for small $Q$, where the oscillations are no longer strictly periodic. This is demonstrated in Fig.~\ref{Fig:Log_gap_q50}.

In addition, as we'll show below, the gapped phases alternate between NI and a symmetry-protected topological phase, denoted LCI$'$, as shown in Fig.~\ref{Fig:Log_gap_q50}.
We expect the gap closing points between these phases to broaden into finite regions due to the finite bandwidth of the LLs, as observed for $\gamma<0$. However, in this case, each transition exhibits a richer structure, with two WSM regions and an LCI phase in between, as shown in Fig.~\ref{Fig:Gap_q5_10}.

\begin{figure}
\includegraphics[width=1\linewidth]{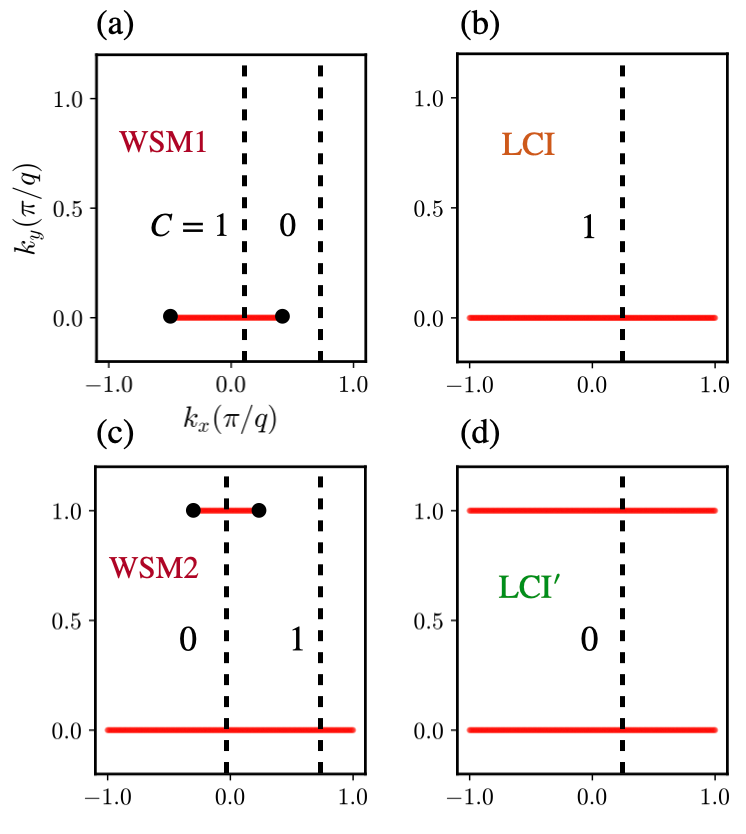}
\caption{Evolution of zero energy surface states in the $k_x$-$k_y$ 
surface BZ in the transitions from an NI to an LCI$'$ state (the transition 
is depicted in Fig. \ref{Fig:Gap_q5_10}(c)) for a fixed flux $1/5$ and 
$\gamma=0.96$. The $Q$ values are 2.175, 2.195, 2.22 and 2.30  in
(a)-(d) respectively.  Figure (a) shows the zero energy Fermi-arc states of the WSM 
state which appear in the  transition NI to LCI. Chern numbers are calculated 
for a fixed $k_x$ value, indicated as vertical dashed lines representing ($k_y,k_z$) 
planes over which we integrate the Berry curvature. (b) As $Q$ is increased, the 
Weyl nodes approach the edge 
of the magnetic BZ, merge, and get annihilated. At the end of this process, 
the corresponding Fermi arc spans the entire BZ along $k_x$. Thus an LCI 
state emerges. (c) A pair of Weyl nodes appear, accompanied by a new Fermi 
arc. This Fermi arc coexists with the surface states of the LCI bands. (d) 
As $Q$ is increased further, the Weyl nodes 
approach the edge of the magnetic BZ, merge at $(\pi/q, \pi/q)$,
and get annihilated. Again, the corresponding Fermi arc now spans the 
entire BZ along $k_x$. Thus an LCI$'$ state emerges. In this state $C(k_x)=0$ 
for all $k_x$.}  
\label{Fig:Surface_NonMono}
\end{figure}

To understand the transition from NI to LCI$'$, let us begin from the NI phase at small $Q$.  As we approach the transition with increasing $Q$, two LL bands approach each other (panel (a) of Fig.~\ref{Fig:Band_Evolution}).  We enter a WSM phase when the two bands touch at the $\Gamma$ point of the magnetic BZ.  Then, two WNs form and they move away from each other as we keep increasing $Q$ (panels (b) and (c)).  When the WNs reach the edge of the magnetic BZ, we enter an LCI phase (panel (d)).  Increasing $Q$ further, we enter another WSM phase when the two bands touch again, but this time at $(k_x,k_y)=(0,\pi/q)$.  These WNs move away from each other (panels (f) and (g)) until they reach the corner of the magnetic BZ, when the system enters the LCI$'$ phase (panel (h)).  This entire sequence is reversed at the next gap closing, when the system evolves from LCI$'$ to NI.

Note that the appearance of the LCI phase requires the gap at $(0,\pi/q)$ to be larger than that at $(\pi/q,0)$.  For the model we consider, this is always satisfied for $\gamma>0$.  This order of the gaps guarantees that the pair of WNs at $k_y=0$ leaves the zone edge before the pair at $k_y=\pi/q$ is created.  If the order were reversed, then the two pairs would coexist and a single WSM phase would separate the NI and LCI$'$ phases.

The evolution of the Fermi arcs on a slab geometry is shown in Fig.~\ref{Fig:Surface_NonMono}.  There, we see that the LCI phase is characterized by a closed Fermi arc that extends across the magnetic BZ.  By contrast, the LCI$'$ phase contains two such Fermi arcs, one along $k_y=0$ and the other along $k_y=\pi/q$.  These arcs have opposite chirality, such that the layer Chern number is zero. However, the LCI$'$ is protected by mirror symmetry, which ensures that the Fermi arcs reside at the mirror symmetric values $k_y=0$ or $k_y=\pi/q$, thus preventing them from merging and forming an NI.  In the absence of mirror symmetry, there is no topological distinction between LCI$'$ and NI.  This is consistent with the observation made earlier, that gap closings are only expected in mirror symmetric models, due to the exact destructive interference between tunneling paths.

\vspace{0.5cm}
\section{Summary and Conclusion}
\label{Sec:DC}

An external orbital magnetic field, which is perpendicular to the direction 
of separation of Weyl cones of opposite chirality, introduces tunneling of 
electrons between the Weyl cones. 
This leads to a hybridization between the cones (note that such a hybridization is allowed since translation symmetry is broken by the presence of the orbital magnteic field) and an opening of a gap.
Such a tunneling process requires overcoming the energy barrier separating the two Weyl nodes, leading to a nonperturbative dependence of the gap on the magnetic field.
The induced gap is significant when the inverse magnetic length is comparable 
to or larger than the momentum-space separation between the Weyl nodes.

Anisotropy in the Weyl cone dispersion, characterized by $\gamma$ as defined in Eq.~(\ref{Eq:Anisotropy_pm}),  plays a crucial role in determining 
the behavior of the induced gap. When $\gamma < 0$, the system enters either a 
normal insulator or an LCI phase, with 
the induced gap increasing monotonically as the magnetic field strength 
grows. In contrast, for $\gamma > 0$ and small Weyl node separation 
$(Q \ll \pi)$, the gap exhibits periodic closings as the field 
increases, causing the system to alternate between a normal insulator 
and a gapped phase labeled LCI$’$, which can be viewed as two copies of LCI with opposite Chern numbers. For $\gamma>0$  and large node separation, the system enters directly into an LCI state.

Throughout, we assumed that the magnetic field is applied along the $z$ axis.  Then, the anisotropy parameter $\gamma$ depends on the ratio of the velocity 
components of the Weyl fermions $v_y/v_{x}$. An identical analysis can be carried out for a field applied along the $y$ axis.  Then, the relevant anisotropy parameter depends on the ratio $v_z/v_x$.  For layered materials, where these two ratios can be very different, this opens the possibility to observe $\gamma>0$ and $\gamma<0$ behaviors on the same system by rotating the direction of the magnetic field, and to access the different induced phases, namely LCI, LCI$'$ or a normal
insulator.

We now turn to discuss briefly some of the experimental implications of our results.
The magnetic field-induced gap opening in WSMs has direct consequences on the longitudinal magnetoresistance \cite{Bednik_Syzranov_2020}. Such behavior has been already observed in experiments in the Weyl materials 
TaAs \cite{Zhang_Jia_2017} and TaP \cite{Ramshaw_McDonald_2018}. 
However, longitudinal magnetoresistance does not give a direct indication of the topology of the gapped phases. In contrast, Hall conductivity measurements provide 
a clearer distinction between the gapped phases, namely the normal insulator, LCI, 
and LCI$'$ phases. A time-reversal symmetry-breaking 
Weyl semimetal with two Weyl nodes separated by a momentum space distance $Q$
along $k_x$ exhibits a non-quantized anomalous Hall conductivity of 
$\sigma_{yz} = \frac{e^2}{2\pi h} Q$ per layer along the $x$-direction. 
Upon entering the LCI phase with applied magnetic field, this Hall conductivity becomes 
quantized at $\sigma_{yz} = \frac{e^2}{h}$ per layer. In contrast, transitions to 
either a normal insulator or the LCI$'$ phase result in a vanishing Hall conductivity. 
To further distinguish between normal insulator and LCI$'$, one may rely on surface-sensitive probes, such as STM, that can detect the presence of the zero-energy modes arising from 
the closed Fermi arc states in LCI$'$.

We briefly comment on the effects of disorder, finite temperature, and interactions 
on the phase diagram. In the presence of disorder, the LCI$'$ phase cannot
be sharply distinguished from a normal insulator, since the crystalline symmetry 
protecting the LCI$'$ phase is generically broken. Moreover, features associated 
with exponentially small gaps and with very narrow regions in parameter space 
are expected to be smeared out by disorder and thermal fluctuations. Nevertheless, 
the phase transitions from the WSM to either a normal insulating phase or an 
LCI phase are expected to remain robust, provided that disorder, temperature, 
and interaction scales are small compared to the magnetic-field-induced gap.

Finally, we comment on orbital magnetic fields that are not perfectly perpendicular 
to the separation of the Weyl nodes.  The momentum parallel to the magnetic field, 
$k_{\parallel B}$, is a good quantum number.  When the field has a component parallel 
to the separation between Weyl nodes, the value of $k_{\parallel B}$ differs at the 
two Weyl nodes, and they no longer hybridize directly.  Hence, even a small misalignment 
of the field relative to the perpendicular direction can drastically suppress the gap opening \cite{Bednik_Syzranov_2020}. However, this effect is mitigated in systems with nodal 
loops\cite{Fang_Topological2015, Bian_Topological_2016, Bzdusek2016, Neupane_Observation_2016, 
Lou2018, Belopolski2019, Abdulla_Internal_2025, Chen2021}.  In such systems, magnetic fields 
generically lead to direct hybridization of nodal states, regardless of the magnetic field orientation.  
This will be discussed in detail elsewhere.

\begin{acknowledgements} 

F.A. thanks the Helen Diller Quantum centre for financial support. 
A.K. acknowledges funding by the Israeli Council for Higher Education 
support program and by the Israel Science Foundation (Grant No. 2443/22).  
D.P. acknowledges support from the Israel Science Foundation (ISF) Grant 
no. 2005/23.

\end{acknowledgements}

\bibliography{wsm}

\end{document}